\documentclass[apj]{emulateapj}

\usepackage{natbib}
\usepackage{graphicx}
\usepackage{apjfonts}

\slugcomment{Submitted for publication in The Astrophysical Journal}

\shorttitle{A Recollimation Shock 80 mas from the Core in the Jet of the Radio Galaxy \object{3C~120}}
\shortauthors{Agudo, G\'omez, Casadio, Cawthorne, \& Roca-Sogorb}

\begin{document}

\title{A Recollimation Shock 80 mas from the Core in the Jet of the Radio Galaxy \object{3C~120}: Observational Evidence and Modeling}

\author{Iv\'{a}n Agudo\altaffilmark{1,2}, 
            Jos\'{e} L.~G\'omez\altaffilmark{1}, 
            Carolina Casadio\altaffilmark{1}, 
            Timothy V. Cawthorne\altaffilmark{3},
            Mar Roca-Sogorb\altaffilmark{1}
            }


\altaffiltext{1}{Instituto de Astrof\'isica de Andaluc\'ia, CSIC, Apartado 3004, 18080, Granada, Spain; jlgomez@iaa.es}

\altaffiltext{2}{Institute for Astrophysical Research, Boston University, 725 Commonwealth Avenue, Boston, MA 02215, USA}

\altaffiltext{3}{School of Computing, Engineering and Physical Science, University of Central Lancashire, Preston PR1 2HE, UK}

\begin{abstract}
We present Very Long Baseline Array observations of the radio galaxy \object{3C~120} at 5, 8, 12, and 15\,GHz designed to study a peculiar stationary jet feature (hereafter C80) located $\sim80$\,mas from the core, which was previously shown to display a brightness temperature $\sim600$ times lager than expected at such distances. The high sensitivity of the images -- obtained between December 2009 and June 2010 -- has revealed that C80 corresponds to the eastern flux density peak of an arc of emission (hereafter A80), downstream of which extends a large ($\sim20$\,mas in size) bubble-like structure that resembles an inverted bow shock. The linearly polarized emission closely follows that of the total intensity in A80, with the electric vector position angle distributed nearly perpendicular to the arc-shaped structure. Despite the stationary nature of C80/A80, superluminal components with speeds up to $3\pm1\,c$ have been detected downstream from its position, resembling the behavior observed in the HST-1 emission complex in \object{M87}. The total and polarized emission of the C80/A80 structure, its lack of motion, and brightness temperature excess are best reproduced by a model based on synchrotron emission from a conical shock with cone opening angle $\eta=10^{\circ}$, jet viewing angle $\theta=16^{\circ}$, a completely tangled upstream magnetic field, and upstream Lorentz factor $\gamma_u=8.4$. The good agreement between our observations and numerical modeling leads us to conclude that the peculiar feature associated with C80/A80 corresponds to a conical recollimation shock in the jet of \object{3C~120} located at a de-projected distance of $\sim$190\,pc downstream from the nucleus.
\end{abstract}


\section{Introduction}
\label{intr}

The radio galaxy \object{3C~120} is a very active and powerful emitter of radiation at all observed wavebands. While classified as a Seyfert 1 \citep{1967ApJ...149L..51B}, its spectrum shows broad emission lines and a complex optical morphology which has been interpreted as the result of a past merger \citep{GarciaLorenzo:2005p178}. Variability in the emission lines has allowed reverberation mapping to yield a black hole mass of $\sim 3\times10^7 M_{\sun}$ \citep{1999ApJ...526..579W}. In X-rays, it is the brightest broad-line radio galaxy, with the X-ray spectrum becoming softer when the intensity increases \citep{1991ApJ...368..138M}, suggesting that the soft X-rays are dominated by the disk instead of the beamed jet. This has made it possible to establish a clear connection between the accretion disk and the radio jet through coordinated X-ray and radio observations \citep{2002Natur.417..625M,Chatterjee:2009p17109}. Along with \object{3C~111}, \object{3C~120} is one of only  two broad-line radio galaxies detected by Fermi-LAT in the GeV photon energy range, such emission being most probably the result of beamed radiation from the relativistic jet observed at intermediate viewing angles \citep{Kataoka:2011p17122}.
    
At radio wavelengths, \object{3C~120} has a blazar-like one-sided superluminal radio jet extending up to hundreds of kiloparsecs \citep{1987ApJ...316..546W,1988ApJ...335..668W,1991MNRAS.251...54M}. Due to its proximity ($z$=0.033), and to its bright millimeter emission \citep{Agudo:2010p12104}, Very Long Baseline Array (VLBA) observations at high frequencies (22, 43, and 86\,GHz) show a very rich inner jet structure containing multiple superluminal components with apparent velocities up to $6\,c$ \citep{1998ApJ...499..221G,Gomez:1999p17133,2001ApJ...549..840H,Jorstad:2005p264,Marscher:2007p213}. Continued monthly monitoring with the VLBA at 22 and 43\,GHz has revealed rapid changes in the total and linearly polarized intensity, interpreted as resulting from the interaction of the jet components with the external medium \citep{2000Sci...289.2317G,Gomez:2001p201}. Longer wavelength Very Long Baseline Interferometry (VLBI) observations have shown evidence for the existence of an underlying helical jet structure \citep{2001ApJ...556..756W}, which has been interpreted as the result of jet precession \citep{Hardee:2005p260,2004MNRAS.349.1218C}
   
Combination of the information from a sequence of 12 monthly polarimetric VLBA observations of \object{3C~120} at 15, 22 and 43 GHz has allowed imaging of the linearly polarized emission within the innermost $\sim10$\,mas jet structure, revealing a coherent in time Faraday screen and polarization angles \citep{Gomez:2008p30}. Gradients in Faraday rotation and degree of polarization across and along the jet are observed, together with a localized region of high ($\sim6000$ rad m$^{-2}$) Faraday rotation measure (RM) between approximately 3 and 4 mas from the core. The existence of this localized region of high RM, together with the observation of uncorrelated changes in the RM screen and RM-corrected polarization angles, suggest that a significant fraction of the Faraday RM found in \object{3C~120} originates in foreground clouds, rather than in a sheath intimately associated with the emitting jet \citep{Gomez:2011p16108}.
  
VLBA observations of \object{3C~120} in November 2007 revealed a component (hereafter C80) located at 80 mas (which corresponds to a de-projected distance of $\sim$190\,pc for a viewing angle of 16$^{\circ}$) with a brightness temperature $T_b\approx5\times10^9$ K, which is about 600 times higher than expected at such distances from the core \citep{RocaSogorb:2010p11823}. Analysis of previous observations (starting in 1982) shows that this component was not detected at frequencies higher than 5 GHz before April 2007, when was first observed at 15 GHz. After this epoch C80 appears in all images (even at the highest frequencies) at the same location without significant changes in its flux density. \cite{RocaSogorb:2010p11823} conclude that the unusually high $T_b$ of C80 could be explained by a helical shocked jet model -- and perhaps some flow acceleration -- but it seems very unlikely that this corresponds to the usual shock that emerges from the core and travels downstream to the location of C80, requiring some other intrinsic process capable of providing a local amplification in the density of high energy particles and/or magnetic field.

Here we present the results from a new observing program designed to study the nature of the C80 feature. Such study is complemented with semi-dynamical and semi-analytical simulations to help us interpret the structure of C80 and its surrounding jet region. The paper is organized as follows: in Sect.~\ref{obs} we describe the new VLBA observations presented here and their data reduction procedures; in Sect.~\ref{res} we report on the results that are obtained directly from the observations; we present the numerical models used to reproduce these observations in Sect.~\ref{sim}. Finally, in Sect.~\ref{sumconcl} we summarize our main results and draw the conclusions from this work.

At the redshift of \object{3C~120}, and under the standard $\Lambda$CDM cosmology (with $H_0$=71 km s$^{-1}$ Mpc$^{-1}$, $\Omega_M=0.27$, and $\Omega_\Lambda=0.73$) that we assume in this work, an angular separation of 1\,mas in the sky corresponds to a projected linear distance of 0.65\,pc, and a proper motion of $1$\,mas/yr is translated to a superluminal speed $2.19$ times larger than the speed of light.

\section{Observations}
\label{obs}
\begin{figure*}
   \centering
   \epsscale{1.0}
   \plotone{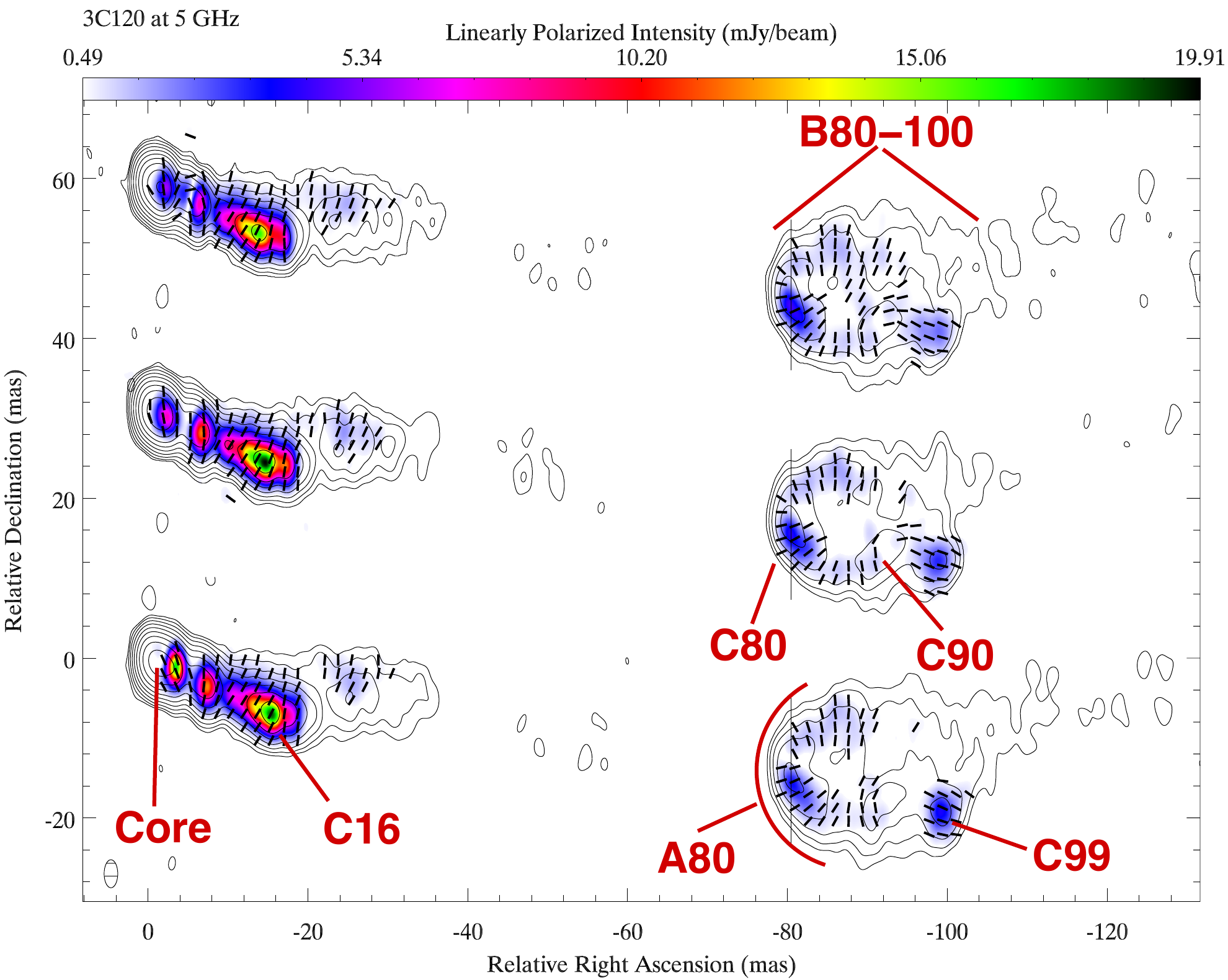}
   \caption{Sequence of VLBA images of 3C~120 at 5\,GHz taken on 2009 December 14, 2010 March 14, and 2010 June 21 (from top to bottom). Vertical separations are proportional to the time difference between epochs. Total intensity contours are overlaid at 0.03, 0.07, 0.17, 0.41, 1.02, 2.49, 6.11, 14.98. 36.72, and 90\,\% of the total intensity peak at 1.53\,Jy/beam. A common convolving beam of FWHM 3.5$\times$1.7 mas at $0^{\circ}$ was used for all images and is shown in the lower left corner. Grey scale images (on a linear scale shown at the top) show the linearly polarized intensity. Black sticks (or unit length) indicate the polarization electric vector position angle, uncorrected for Faraday rotation. The vertical lines at $\sim80$\,mas from the core indicate the location where the transverse cuts of the fractional polarization of the jet shown in Fig.~\ref{pcut} were performed. (A color version of this figure is available in the online journal.)}
   \label{5ghz}
\end{figure*}

\begin{figure*}
   \centering
   \epsscale{1.0}
   \plotone{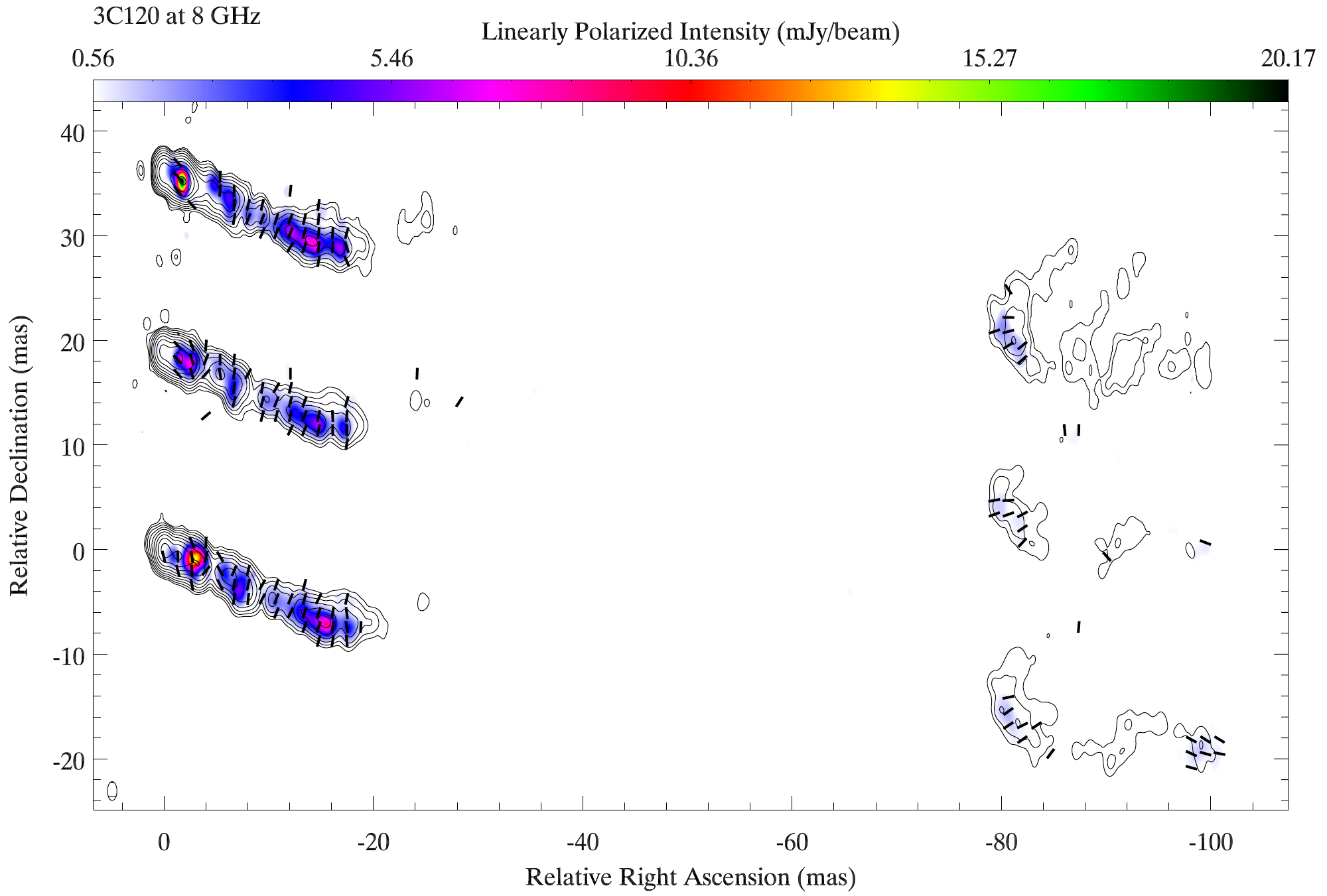}
   \caption{Same as Fig.~\ref{5ghz} but for the 8\,GHz images. Contours levels are drawn at 0.11, 0.23, 0.49, 1.03, 2.17, 4.58, 9.64. 20.30, 42.74 and 90\,\% of the total intensity peak at 1.04\,Jy/beam. A common convolving beam of FWHM 1.8$\times$0.9 mas at $0^{\circ}$ was used for all images. (A color version of this figure is available in the online journal.)}
   \label{8ghz}
\end{figure*}

\begin{figure*}
   \centering
   \epsscale{1.0}
   \plotone{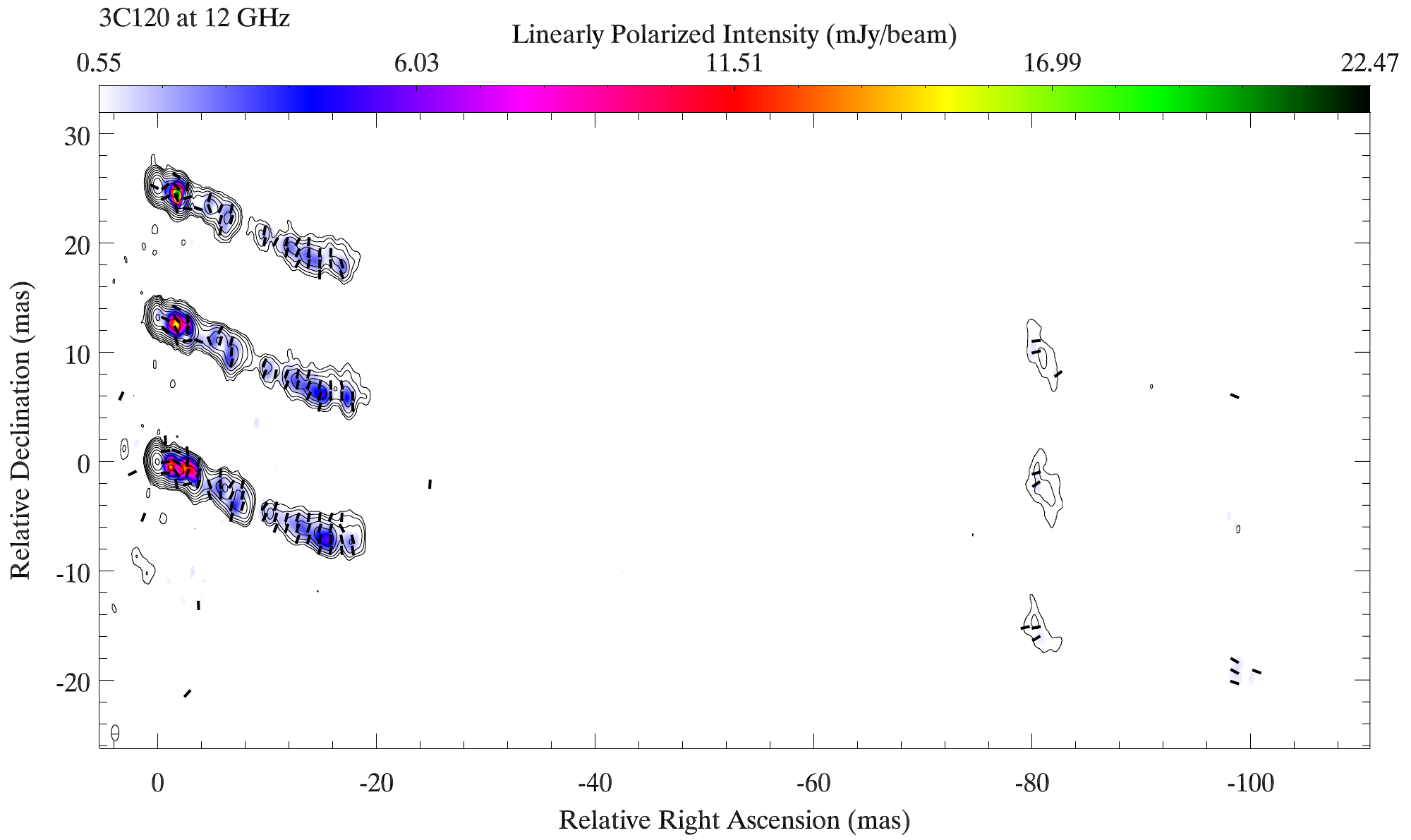}
   \caption{Same as Fig.~\ref{5ghz} but for the 12\,GHz images. Contours levels are drawn at 0.17, 0.34, 0.69, 1.38, 2.76, 5.54, 11.13. 22.34, 44.83 and 90\,\% of the total intensity peak at 0.75\,Jy/beam. A common convolving beam of FWHM 1.5$\times$0.7 mas at $0^{\circ}$ was used for all images. (A color version of this figure is available in the online journal.)}
   \label{12ghz}
\end{figure*}

The new multi-frequency and polarimetric VLBA images of the jet in \object{3C~120} presented in this paper were obtained from observations performed on 2009 December 14, 2010 March 14, and 2010 June 21. The corresponding observations, that employed all ten antennas of the VLBA, were performed at 5, 8, and 12\,GHz (see Figs.~\ref{5ghz}--\ref{12ghz}), and used a 2-bit signal sampling to record in a 32 MHz bandwidth per circular polarization at a recording rate of 256 Mbits/s.
  
Calibration of the data was performed within the AIPS software package following the standard procedure for VLBI polarimetric observations \citep{1995AJ....110.2479L}. To correct for opacity effects on the higher frequency observations ($>8$\,GHz) we used the recorded variation of the system temperature on every station to solve for the receiver temperature and zenith opacity. The final images were obtained following the standard hybrid--mapping procedure through an iterative process that employed both AIPS and the DIFMAP package \citep{Shepherd:1997p17060}.

For the calibration of the absolute orientation of the electric vector position angle (EVPA) we employed both, a set of VLA observations of our polarization calibrators (\object{NRAO150}, \object{OJ287}, \object{3C~279}, and \object{3C~454.3}) from the VLA/VLBA Polarization Calibration Program\footnote{\tt http://www.vla.nrao.edu/astro/calib/polar/}, and the information extracted by the comparison of the D-terms at different observing epochs \citep[see][]{2002.VLBA.SM.30}. The useful input from the first of these two methods was rather limited in our case because the VLA/VLBA Polarization Calibration Program does not include data at 12\,GHz, and the program was halted at the end of 2009 to focus on EVLA commissioning, hence preventing us from getting the calibration at 5 and 8\,GHz for the latest two observing epochs in 2010. However, we verified that the 5 and 8\,GHz D-terms were very stable during our three observations. \citet{2002.VLBA.SM.30} showed that in this case the D-terms can be used to calibrate those observing epochs without an independent EVPA calibration. This allowed us to calibrate the 5 and 8\,GHz EVPA of the latest two observing epochs by rotating their corresponding D-terms to match those of the fully--calibrated first observing epoch.

\begin{figure*}[t]
   \centering
   \epsscale{1.0}
   \plotone{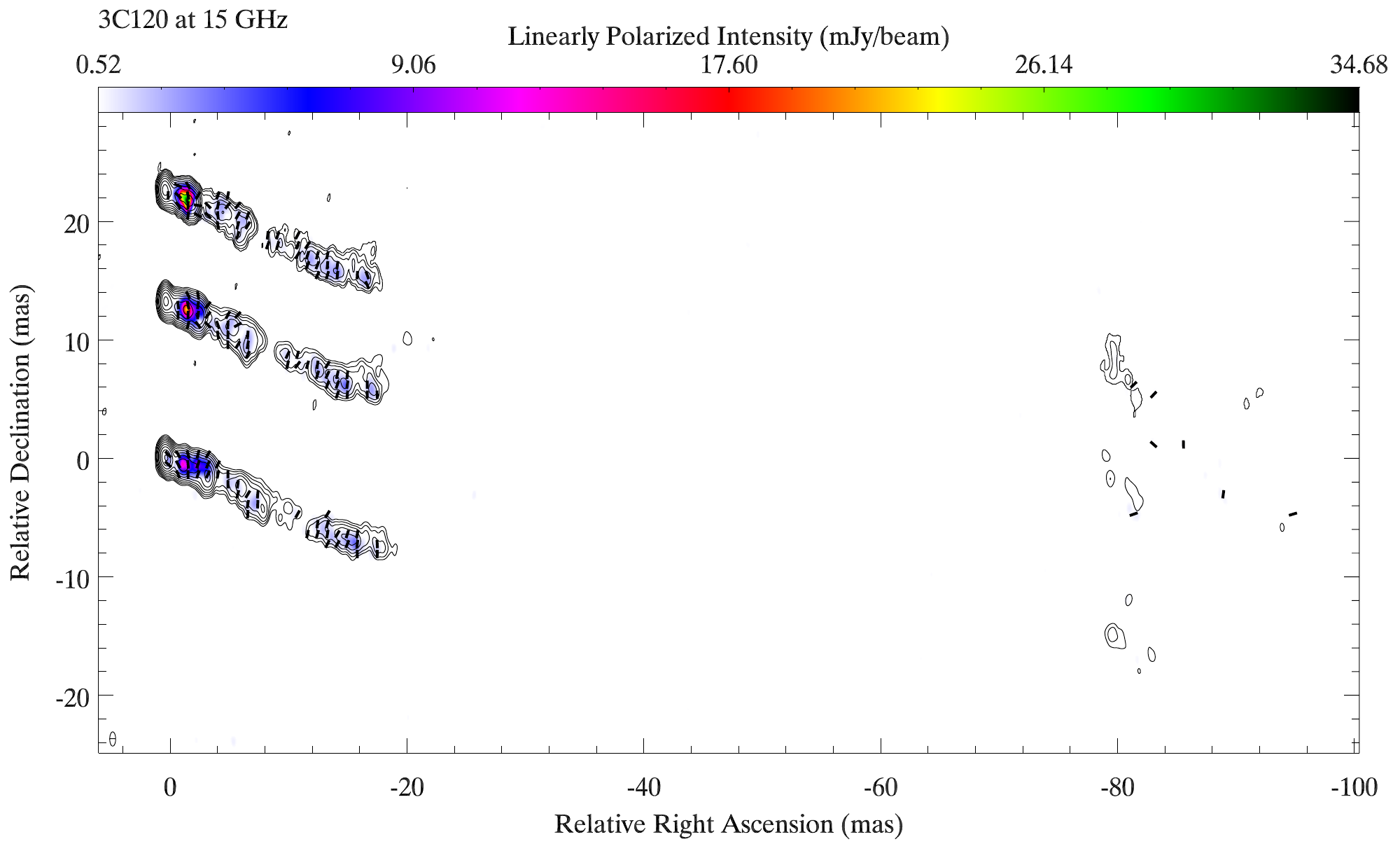}
   \caption{Same as Fig.~\ref{5ghz} but for the 15\,GHz images obtained and calibrated by the MOJAVE team (see the text). From top to bottom, the three images correspond to VLBA observations performed on 2009 December 10, 2010 March 10, and 2010 July 12. Contours levels are drawn at 0.13, 0.27, 0.56, 1.15, 2.38, 4.92, 10.17. 21.04, 43.52 and 90\,\% of the total intensity peak at 0.73\,Jy/beam. A common convolving beam of FWHM 1.2$\times$0.5 mas at $0^{\circ}$ was used for all images. (A color version of this figure is available in the online journal.)}
   \label{15ghz}
\end{figure*}

At 12\,GHz there are no available independent polarization data from our sources to estimate the adequate EVPA calibration. However, we used the publicly--available total--intensity and linear--polarization 15\,GHz VLBA images of \object{3C~120} taken (and calibrated) by the MOJAVE team\footnote{\tt http://www.physics.purdue.edu/MOJAVE/} on dates close to those of our multi-frequency observations (see Fig.~\ref{15ghz}). These 15\,GHz EVPA images allowed us to calibrate our 12\,GHz EVPA on nearby epochs by fitting Faraday rotation profiles at 5, 8 and 15\,GHz to find the correct EVPA calibration at 12\,GHz. The stability of the D-terms at 12\,GHz was also used as a cross-check for the final calibration at this frequency. The final errors in our absolute calibration of the EVPA at all our observing frequencies are estimated to be $\sim5^{\circ}$.

\section{Results}
\label{res}

\subsection{The Inner $\sim20$\,mas of the Jet}
\label{innerjet}
Figures~\ref{5ghz} to \ref{15ghz} show the new 5, 8, 12, and 15\,GHz VLBA images of \object{3C~120} in total and linearly polarized intensity, as well as their corresponding EVPA distribution. These new images show a straight jet within the inner $\sim20$\,mas that contains a bright and highly polarized knot at $\sim16$\,mas from the core (C16). The EVPA of C16 is distributed perpendicular to the jet axis. The same EVPA orientation is found in the C16 counterparts at 8, 12, and 15\,GHz, see Figs.~\ref{8ghz}-\ref{15ghz}, where the images show an increasingly rich structure of jet knots with observing frequency all along the innermost $20$\,mas of the jet. The multi-frequency polarization properties of C16 match those of the C12 jet feature reported by \citet{Gomez:2011p16108}. This, together with the total flux dominance of both C16 and C12 in regions $\gtrsim10$\,mas from the core and the fact that C16 lies in the jet region expected from the position of C12 and the typical proper motions in the jet of \object{3C~120} \citep[$\sim2$\,mas/yr,][]{Gomez:2001p201}, allow us to identify both components as representing the same jet moving knot.

\subsection{Bright Emission Region Between 80 and 100\,mas from the Core: B80-100}
\label{B80-100}

The jet feature at $\sim80$\,mas from the core (C80) is detected in all new images presented in this work from 5 to 15\,GHz (Figs.~\ref{5ghz}-\ref{15ghz}). No other jet feature is detected in the jet downstream of C80 in the 12 and 15\,GHz images presented here. However, the new 5\,GHz images show an emission region in an arc from north to south around the eastern side of C80, having the shape of an inverted bow shock, similar to those observed at the end of the large scale jets of FR-II radio galaxies \citep[e.g.,][]{Perley:1984p17139}. Perhaps because of the lack of sensitivity, this arc of emission (that we have labeled as A80) was not detected by our previous short-integration VLBA observations of the source at 5\,GHz \citep[see][]{RocaSogorb:2010p11823,Gomez:2011p16108}, where only the detection of C80, the brightest region in A80, was reported.

Together with C80/A80, the new 5\,GHz images show a bubble-like extended emission--region larger than $\sim20$\,mas along the jet axis -- from the sharp edge near the location of C80--, and $\sim20$\,mas across the jet axis. This emission structure will be called B80-100 hereafter (see Fig.~\ref{5ghz}). Neither the total intensity nor the polarized emission is symmetric with regard to the jet axis in B80-100. The maxima in total and linearly polarized intensity are both located on the southern side of the bubble, which resembles the behavior of the jet in \object{3C~120} within the first $\sim10$\,mas \citep{2000Sci...289.2317G,Gomez:2001p201,Gomez:2008p30}.

Further downstream from C80 (but still within B80-100), C90, a more extended jet region reported by \citet{RocaSogorb:2010p11823} and \citet{Gomez:2011p16108} at observing frequencies below 8\,GHz, starts to be resolved in our new images (Figs.~\ref{5ghz} and \ref{8ghz}). Our 8\,GHz images in 2009 and 2010 show the jet emission around C90 as a slightly more extended and sparser jet region than in the image taken in 2007 November 30 by \citet{Gomez:2011p16108} at the same frequency. C90 is better observed in our 5\,GHz images, that map the entire B80-100 at higher sensitivity. This provides a reliable superluminal speed measurement of C90 ($v_{\rm{C90}}=3.4\pm1.0\,c$), which contrasts with the stationary character of C80 \citep{RocaSogorb:2010p11823} located $\sim10$\,mas upstream in the jet.

Our new images also show a bright and compact jet region located $\sim99$\,mas from the core -- reported here for the first time -- that we have labeled C99. Our kinematic study of this feature also reveals superluminal proper motion ($v_{\rm{C99}}=3.0\pm1.1\,c$) of similar magnitude and direction downstream the jet than C90. The superluminal character of the flow downstream from C80 resembles the behavior near the HST-1 knot in \object{M87} as reported by \cite{Cheung:2007p685}, who suggested HST-1 as the site of a recollimation shock in the jet of \object{M87} \citep[see also][]{Stawarz:2006p17247,Asada:2012p20001}.

\begin{figure}
   \centering
   \epsscale{1.1}
   \plotone{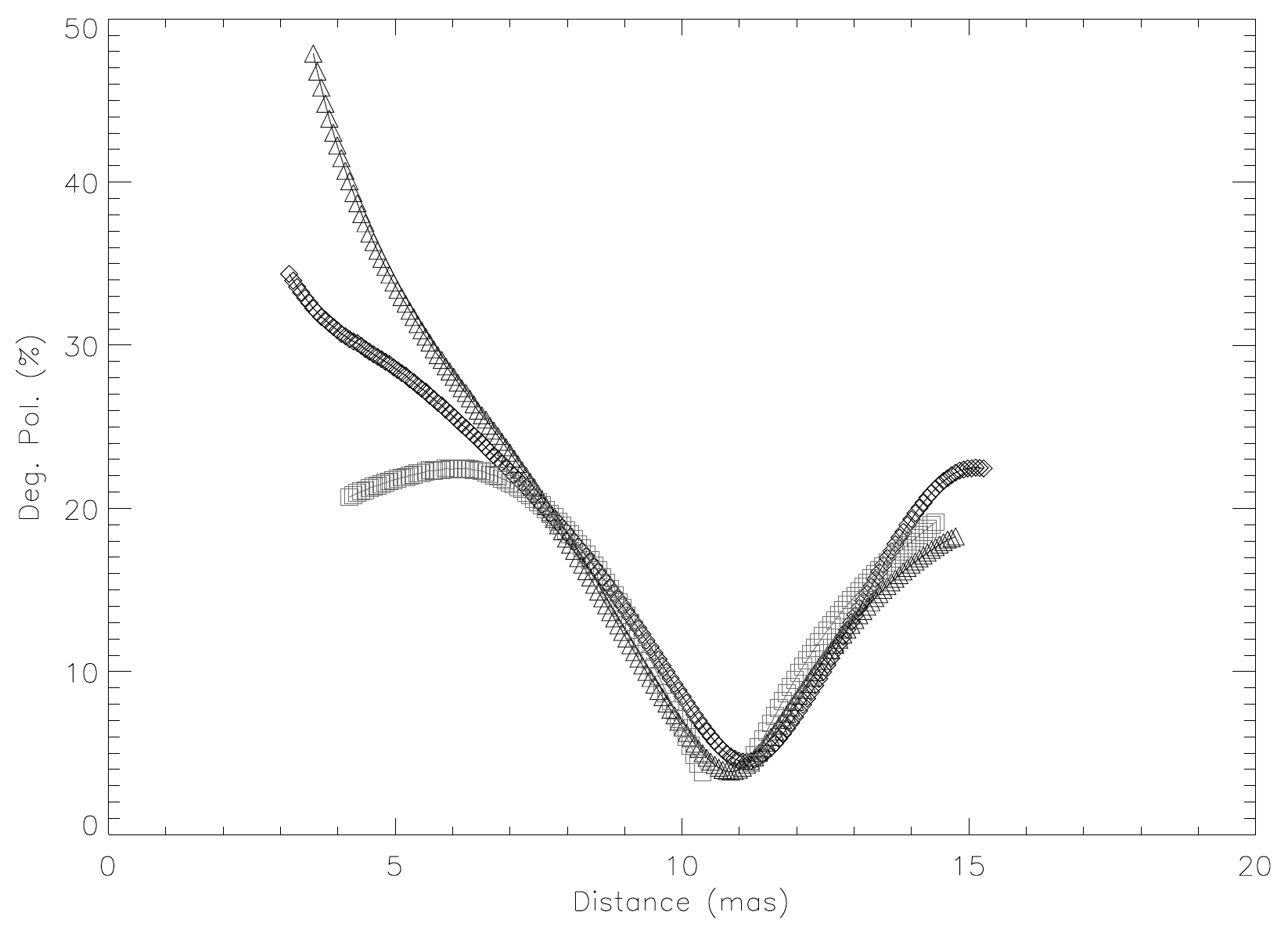}
   \caption{5\,GHz fractional polarization across the A80 jet structure through the slices (from south to north) marked in Fig.~\ref{5ghz}. Diamonds correspond to epoch 2009 December 14, triangles for epoch 2010 March 14, and squares for 2010 June 21.}
   \label{pcut}
\end{figure}

\begin{figure*}
   \centering
   \epsscale{1.}
   \plotone{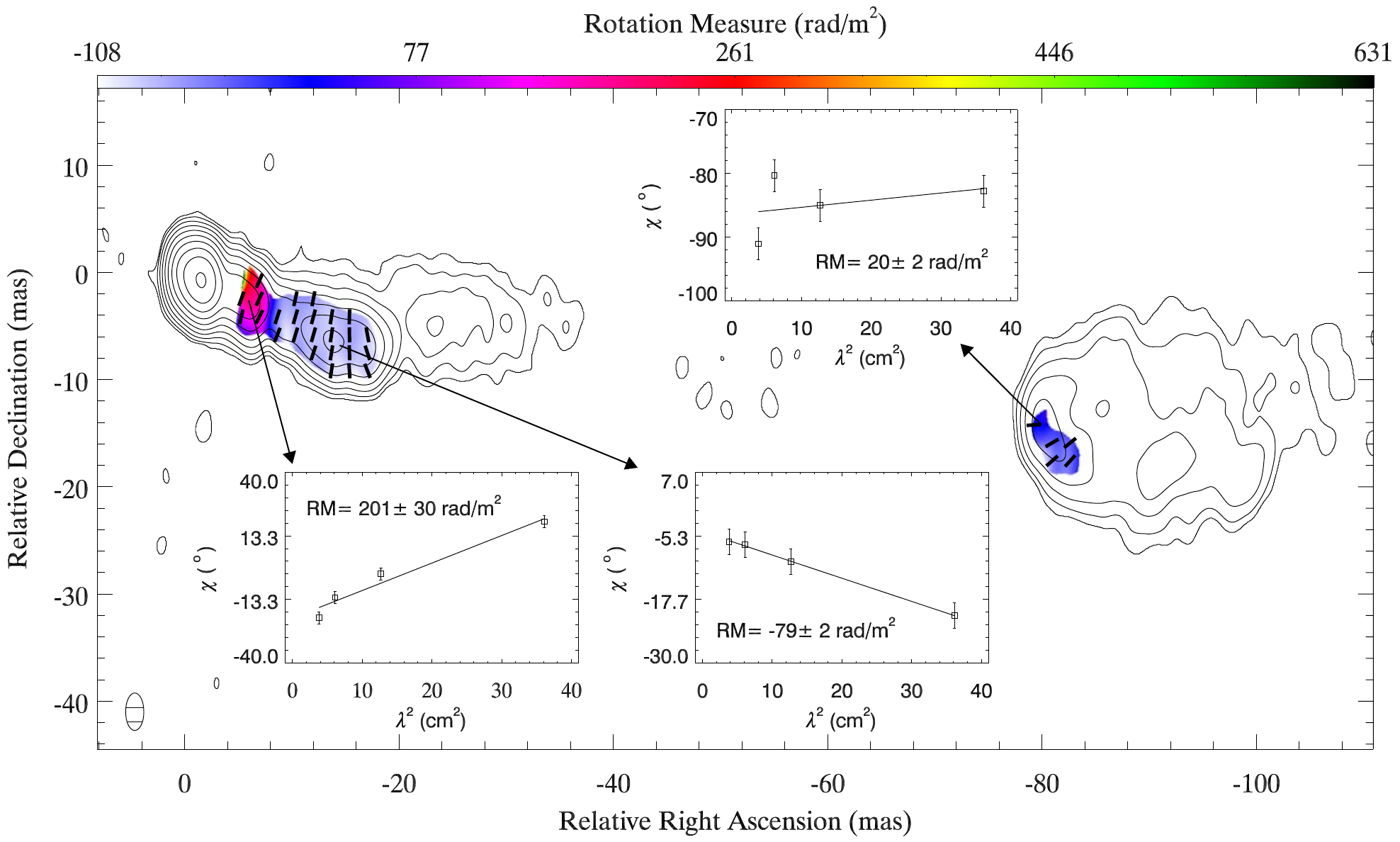}
   \caption{Rotation measure image of 3C~120 between 5 and 15 GHz for observations taken in 2009 December 14. Contours show the 5 GHz total intensity image (see Fig.~\ref{5ghz}). Black sticks indicate the RM-corrected EVPAs. Inset panels show sample fits to a $\lambda^2$ law of the EVPAs at some particular locations. The convolving beam is shown in the lower left corner. (A color version of this figure is available in the online journal.)}
   \label{rm}
\end{figure*}

Given the typical superluminal motions previously detected in \object{3C~120} on smaller jet scales \citep[$\sim4\,c$;][]{Gomez:2001p201}, we consider it unlikely that C90 -- as identified by \citet{RocaSogorb:2010p11823} and \citet{Gomez:2011p16108} -- can be identified with C99 at earlier epochs, given the exceedingly large proper motion that would be needed ($\gtrsim9\,c$, which has not previously been observed in the jet of \object{3C~120}). Our 5\,GHz images show that C99's flux density increases progressively from 2009 December 14, to 2010 June 21, which provides a better explanation for why C99 was not detected in our previous 5\,GHz observations in 2007 November 30.

\subsection{Arc of Emission at $\sim80$\,mas from the Core: A80}
\label{arc}
  
As it is shown in Fig.~\ref{5ghz} (see also Figs.~\ref{8ghz} to \ref{15ghz}), the linearly polarized emission in A80 closely follows that of the total intensity, with values of the degree of polarization up to $\sim30$\% (see also Fig.~\ref{pcut}), and the EVPAs distributed nearly perpendicular to the arc-shaped structure of A80, as expected in the case of a shock front.

The observations presented here show a nearly constant flux density evolution of C80/A80, and confirm the large brightness temperature as well as its stationarity, both within our new observing epochs and with regard to the reported position of C80 at earlier epochs. These properties appear remarkably similar to those expected in the case of conical recollimation shocks \citep[e.g.,][]{Cawthorne:1990p17114,1995ApJ...449L..19G,1997ApJ...482L..33G,Agudo:2001p460,Cawthorne:2006p409, Nalewajko:2009p17298}, and will be discussed in detail in the next session. 

Interestingly, the region of B80-100 close to the jet axis and downstream from the sharp edge of A80 appears unpolarized in our 5\,GHz images. In contrast, weak linear polarization emission (with EVPA nearly perpendicular to the local jet axis) starts to be discerned further downstream at $\sim10$\,mas from the eastern edge of A80. The strong linear polarization emission of C99 shows its EVPA distribution parallel to the jet axis, which, together with the measured superluminal motion, may be an indication of a travelling plane shock wave perpendicular to the axis.

\subsection{Faraday Rotation Screen Between 5 and 15\,GHz}
\label{rms}

  Figure \ref{rm} shows the rotation measure image of 3C~120 for December 2009 made by combining the data between 5 and 15\,GHz. The RM images for the other two epochs of our observations show very similar results as those for December 2009, and are therefore not shown. As can be seen in the inset panels of Fig.~\ref{rm}, we obtain good fits to a $\lambda^2$ law of the EVPAs throughout the jet, except at longer wavelengths in the innermost $\sim$4 mas, probably due to beam depolarization and opacity.

  The RM screen shown in Fig.~\ref{rm} is in excellent agreement with that obtained for observations taken in November 2007 \citep[see Fig.~7 in][]{Gomez:2011p16108}, providing further confirmation for the stationarity of the Faraday screen in 3C~120 claimed previously \citep{Gomez:2008p30}. Note, however, that relatively rapid changes (in scales of tens of months) in the RM have also been observed in the innermost $\sim$2 mas of the jet \citep{Gomez:2011p16108}. Our new observations also confirm the negative gradient in RM with distance along the jet, including a sign reversal at $\sim$8 mas from the core. The motion of component C16 along the jet during the two years time span between the 2007 observations of \citet{Gomez:2011p16108} and ours have allowed the mapping of the RM further downstream, revealing a RM of $-79\pm2$ rad m$^{-2}$, fully consistent with the RM gradient along the jet mentioned previously.
  
  Component C80/A80 has a small rotation measure of $20\pm2$ rad m$^{-2}$, leading to a Faraday rotation in the EVPAs at our longest observing wavelength of 6 cm (5\,GHz) of 4 degrees, within the estimated error in our absolute calibration of the EVPAs. We therefore can conclude that our EVPA maps of C80/A80 are not affected by Faraday rotation.
  
\section{A conical shock model for A80.}
\label{sim}

The structure and polarization of the A80 feature point towards an origin in the kind of shock that can occur when the internal pressure of a jet adjusts as a result of changes to its environment. This type of structure has been suggested as the origin of the HST-1 knot in M87 \citep{Stawarz:2006p17247,Asada:2012p20001}. A jet that is initially freely expanding may become underpressured and is therefore recollimated by a conical shock converging away from the nucleus. Such shocks may reflect from the jet axis, producing a second conical shock diverging away from the nucleus, that further adjusts the collimation of the jet. The result is a structure consisting of a pair of point-to-point conical shocks that have been found to occur both in experiments and numerical simulations \citep[e.g.,][]{1995ApJ...449L..19G}. In A80, there is no evidence for a converging shock --the shape of the structure, particularly in polarized intensity, suggests a diverging cone that will divert the flow away from the axis causing a drop in pressure. This therefore appears to indicate that A80 has arisen because the jet has rapidly become overpressured, perhaps as a result of a sudden drop in external pressure at this point. Though this seems the most probable explanation it is possible that the jet has encountered small object -- perhaps a molecular cloud typical of the narrow-line emission regions in AGN --  near the axis of the jet, and that the conical shock represents the resulting wake. This explanation seems less probable because it depends on the lucky chance that the obstacle is near the jet axis, but it is in agreement with previous indications for a jet/cloud collision at distances closer to the core \citep{2000Sci...289.2317G}.

Here, the A80 feature is modelled as a diverging conical shock using the models described by \citet{Cawthorne:1990p17114} and \citet{Cawthorne:2006p409}. These models assume the plasma to be a relativistic gas with sound speed $c/\sqrt{3}$.  

Previous monitoring of the proper motions in 3C~120 \citep[e.g.,][]{2000Sci...289.2317G,Gomez:2001p201,2001ApJ...556..756W,2001ApJ...549..840H,2002Natur.417..625M,Jorstad:2005p264} shows that the maximum apparent velocity observed is $\sim6\,c$. From this it is possible to estimate the maximum viewing angle, $\theta_{max}=19^{\circ}$, and the minimum Lorentz factor of the jet, $\gamma_{u}=6$. \citep[e.g.,][]{2000Sci...289.2317G}. The value $\beta_{app}=6$ is therefore assumed, being also relevant to the components labelled L5 to L8 by \citet{2001ApJ...556..756W} in the region upstream from C80. This value is used to constrain the properties of the upstream flow. The 5\,GHz images presented in this paper are used to constrain the properties of the shock wave.

The projected semi-opening angle $\eta_{p}$ of the conical shock is best estimated from the polarized intensity images presented earlier in this paper (Fig.~\ref{5ghz}). The cone axis is taken to be a line inclined at approximately $7^{\circ}$ from EW, pointing directly back to the radio core structure. The opening angle deduced from the distribution of polarized flux density is about $90^{\circ}$, but after convolution of model structures with the elliptical beam, the best fit to this is found from a projected opening angle in the region of $80^{\circ}$ corresponding to $\eta_p=40^{\circ}$. The angle between the polarization rods on either side of the structure (measured where the polarized flux density is brightest) is in the region of $90^{\circ}$. Since the model polarization $E$ rods are orthogonal to the projected edge of the conical structure, a value of $\eta_p=40^{\circ}$ corresponds to an angle of $100^{\circ}$ between the polarization rods on the two sides. However, after convolution, this angle is reduced to approximately $90^{\circ}$, in agreement with the images, as a result of the influence of the polarized structure nearer the axis. A value of $\eta_P$ much less than $40^{\circ}$ would be hard to reconcile with both a symmetrical cone structure and an axis pointing back to the core region; if the axis is maintained at its inclination of $7^{\circ}$ to EW, then the southern edge of the cone would then exclude much of the observed flux density. Values of $\eta_p$ significantly greater than $40^{\circ}$ result in simulated polarization images that are inconsistent with those observed in that they do not reproduce the observed saddle point between the two sides of the polarization image. Hence, a value for $\eta_p$ in the region of $40^{\circ}$ seems to offer the best chance of modelling the observed structure of A80, and is therefore used in what follows.

The true semi-opening angle $\eta$ is related to $\eta_{p}$ by

\begin{eqnarray}
   \tan\eta_{p} = \frac{\tan\eta}{\cos\theta(1-\tan^2\theta\tan^2\eta)^{1/2}}
\end{eqnarray}
which can be solved numerically for $\eta$ in terms of $\eta_{p}$ and $\theta$. 

\begin{deluxetable*}{lccccccccccc}
\tablecolumns{12}
\tabletypesize{\scriptsize}
\tablewidth{0pt} 
\tablecaption{\label{T1} Model parameters for different values of $\theta$.}
\tablehead{\colhead{Model} & \colhead{$\theta$} & \colhead{$\beta_u$} & \colhead{$\eta$} & \colhead{$\kappa$} & \colhead{$\xi$} & \colhead{$\beta_d$}  & \colhead{$\delta_u$} & \colhead{$\delta_{d,max}$} & \colhead{$\delta_{d,max}/\delta_{u}$} & \colhead{$R_1$} & \colhead{$R_2$}\\
                 \colhead{ } & \colhead{$^{\circ}$} & \colhead{} & \colhead{$^{\circ}$} & \colhead{} & \colhead{$^{\circ}$} & \colhead{}\\  
                 \colhead{(1)} & \colhead{(2)} & \colhead{(3)} & \colhead{(4)} & \colhead{(5)} & \colhead{(6)} & \colhead{(7)} & \colhead{(8)} & \colhead{(9)} & \colhead{(10)} & \colhead{(11)} & \colhead{(12)}}
\startdata
1            & 12 & 0.9874    & 7.5  & 0.76     & 1.2 & 0.9849  &  4.63  & 4.73     &  1.02 &   2.9   &  3.0   \\  
2            & 14 & 0.9895    & 8.8  & 0.50     & 3.2 & 0.9826  &  3.62  & 5.33     &  1.47 &  40.3   & 27.5  \\
3            & 16 & 0.9929    & 10.0 & 0.31     & 5.0 & 0.9815 &  2.61  & 5.26     &  2.02 & 600   & 395      \\ 
4            & 18 & 0.9974    & 11.2 & 0.14     & 6.9 & 0.9810 &  1.40  & 5.19     &  3.71 & 68,900  & 39,300 
\enddata
\tablecomments{For all these models the upstream magnetic field was assumed to be completely tangled. \\
                          Columns are as follows:        
                         (1) model number, 
                         (2) angle $\theta$ between the axis and the line of sight, 
                         (3) $\beta_u$ upstream flow velocity as a fraction of $c$, 
                         (4) $\eta$, the (deprojected) semi-opening angle, 
                         (5) $\kappa$, the compression factor at the shock front, 
                         (6) $\xi$, the angle through which the flow is deflected, and
                         (7) $\beta_d$, the downstream flow velocity expressed as a fraction of $c$.
                         (8) $\delta_u$ Doppler factor upstream the conical shock, 
                         (9) maximum Doppler factor downstream the conical shock $\delta_{d,max}$, 
                         (10) $\delta_{d,max}/\delta_{u}$ ratio of maximum downstream to upstream Doppler factor, 
                         (11) intensity ratio between the $R_1$ downstream and upstream emission, 
                         and (12) a new $R_2$ estimate of the downstream to upstream intensity with downstream intensity weighted by the Doppler shift. 
                         }
\end{deluxetable*}

A further constraint is provided by the requirement that, in the rest frame of the upstream flow, the shock wave must advance toward the stationary plasma at speed greater than the sound speed, $c/\sqrt{3}$. This requires that

\begin{eqnarray}
   \beta_{u} \gamma_u \sin \eta > 1/\sqrt{2} \label{condition}
\end{eqnarray}

For each value of $\theta$, a unique model can be computed using the procedures described in \citet{Cawthorne:2006p409} provided inequality~\ref{condition} is satisfied. Parameters appropriate to several models are given in Table~\ref{T1} and the corresponding model images have been computed assuming that the upstream magnetic field is completely disordered. The plausible range of $\theta$ is quite restricted: values below about $10^{\circ}$ are ruled out about by Eq. \ref{condition}, while values above $18.9^{\circ}$ cannot yield apparent superluminal speeds as high as $\simeq 6c$.

\begin{figure*}
   \centering
   \epsscale{1.1}
   \plotone{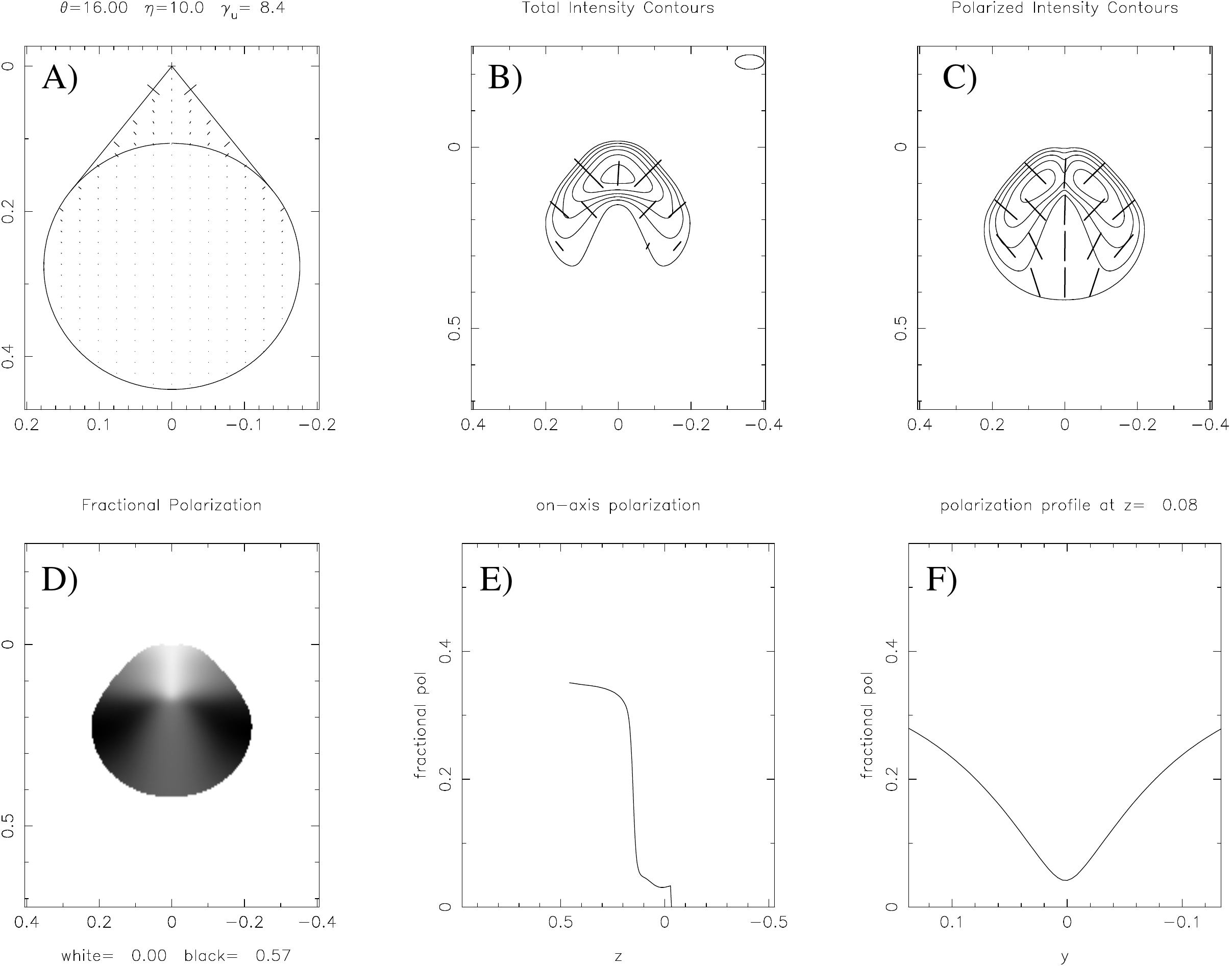}
   \caption{Simulated images and plots corresponding to model 3 (see Table~\ref{T1}). Plots of the six panels are as follows: A) outline of the shock structure and the (unconvolved) polarization rods of length proportional to $P$ , B) contours of $I$ after convolution with a beam chosen to match in shape that of the observations; the convolved polarization is shown by rods parallel to the $E$ field and of length proportional to $P$, C) contours of convolved $P$, with polarization angle $\chi$ shown by the orientation of rods of constant length, D) variation of fractional polarization (from the convolved images), E) along the jet axis, and F) across the jet through the region of the $I$ peak. In all six plots the units of distance are scaled so that the height of the cone (from apex to base) is unity.}
   \label{mod}
\end{figure*}

The simulated images corresponding to models 1--4 resemble the observations of A80 shown in Fig.~\ref{5ghz}. The total intensity ($I$) images in Fig.\ref{mod} show a bow wave pattern while the simulated polarized intensity ($P$) images show a similar pattern that is divided by a saddle point at the position of maximum $I$. The polarization $E$ rods in the two elongated $P$ features are approximately perpendicular to the direction of elongation, which corresponds to the outline of the conical shock wave. Models $1$ to $4$ produce simulations that differ mainly in two ways: first, the structures become more compact as $\theta$ increases, because the emission from the far side of the shock wave becomes relatively weaker; and second, the degree of polarization increases as $\theta$ increases, mainly because the shock becomes stronger (the compression factor of the shock $\kappa$ decreases) as a result of increasing values of $\eta$ and the upstream flow velocity ($\beta_u$). 

The model best matching the degrees of polarization found in the images is model 3. In this model, the degree of polarization rises to about $30\%$ near the edges of A80 in the region of the peak in total intensity, and levels as high as this are seen in this region on the southern side of A80 (as shown in the profiles plot of Fig.~\ref{pcut}).

The simulated images for model 3 are shown in Fig.~\ref{mod}. The model is reasonably successful in reproducing the observed structure in the vicinity of the $I$ peak. However, further downstream the observed polarization rods turn to become orthogonal to the axis whereas the model polarization rods turn in the opposite sense. On axis, the model polarization rods are parallel to the axis, but where on-axis polarization is seen in the observations, at distances greater than about $10$\,mas from the working surface,  the rods are either orthogonal or oblique to the axis. It seems that if the conical shock model is appropriate, then it is the dominant influence on the polarized emission only within about $10$\,mas of the apex. 

Another way in which the model is successful is in its prediction of a very large ratio of upstream to downstream intensity. The observations of \citet{RocaSogorb:2010p11823} have shown that C80 (the brightest hot spot in A80) represents an increase in the intensity of emission with respect to the underlying jet by a factor of approximately $600$, a figure that they found difficult to explain without invoking acceleration of the jet. If this is due to the formation of a shock structure, then the ratio of the up to downstream intensity should be about $600$ or greater. If the Doppler factors of the up and downstream flows are $\delta_u$ and $\delta_d$, respectively, and the compression ratio is $\kappa$, then the intensity ratio should be approximately \[ R_1 = \frac{I_d}{I_u} = \kappa^{(5\alpha-6)/3} \left(\frac{\delta_d}{\delta_u}\right)^{(2-\alpha)} \] where $\alpha\simeq-1$ is the spectral index for C80 \citep{RocaSogorb:2010p11823}. In estimating the value of $R_1$ the largest downstream Doppler factor $\delta_{d,max}$ (for flow at angle $\theta-\xi$ to the line of sight) has been assumed. The values of $\kappa$, $\delta_{d,max}$ for the downstream flow closest to the line of sight and $R_1$ are shown in Table~\ref{T1}. Because $\delta_d < \delta_{d,max}$ for most of the downstream flow, $R_1$ is likely to be an overestimate of the intensity ratio, and so a second estimate based on an intensity weighted average of the downstream Doppler factor, $R_2$, has also been included in the table.

Table~\ref{T1} shows that, as $\theta$ increases from $12^{\circ}$ to $18^{\circ}$, so the downstream to upstream brightness ratios, $R_1$ and $R_2$ increase very rapidly. The greater part of this effect is due to the compression of the plasma (indicated by the value of $\kappa$), and its effect on the magnetic field and particle density. However the increase in the value of $\delta_{d,max}/\delta_u$ is also significant. This occurs because (i) $\delta_u$ is decreasing because $\theta$ is increasing and $\beta$ is also increasing in the regime where $\sin\theta > \gamma_u^{-1}$; and (ii) because $\delta_d$ is varying only weakly, which occurs because although $\beta_d$ decreases and $\theta_d$ increases, the shock becomes stronger and parts of the downstream flow are directed closer to the observer's line of sight.

From Table~\ref{T1} it is clear that parameters similar to those of model 3 ($\theta=16^{\circ}$), which gave the closest match for the fractional polarization values, should have a downstream to upstream brightness ratio of order several hundred. It therefore seems likely that such models can reasonably reproduce both the observed polarization structures and the very high ratio of down to upstream intensity demanded by \citet{RocaSogorb:2010p11823}. 

A few variants on the models described here have been investigated to determine whether they might explain the transverse polarization rods seen on the downstream side of C80. First a poloidal magnetic field component was added, as described by \citet{Cawthorne:2006p409}. The influence of the poloidal field on the polarization is strongest on the near side of the shock front, producing transverse polarization rods near the peak in $I$, where none are seen. The effect is negligible on the far side of the shock, which extends further downstream toward the region where transverse polarization rods would be desirable. The possibility that the upstream flow might be converging slightly has also been investigated, as suggested by \citet{Nalewajko:2009p17298}. The effect was to increase the degree of polarization at a given value of $\theta$, but otherwise the simulated images were very similar. It seems that neither of these modifications helped to explain the larger scale polarization properties beyond A80. It is possible that allowing for a more complex upstream field structure might explain the entire polarization structure of B80-100, but on the other hand, it seems likely that the conical shock is not the dominant influence on polarization toward the downstream edge of the B80-100 structure. Indeed, the superluminal motion found for C90 and C99 suggests that instead the polarization structure downstream of A80 may be the result of moving plane-perpendicular shock waves that appear on the wake of the A80 conical shock.

It is worth noting that, although the upstream Doppler factor is rather low in model 3 ($\simeq 2.6$) the jet to counter-jet ratio is still reasonably high, giving \[ \frac{I_{jet}}{I_{counterjet}} = \left(\frac{1+\beta_u \cos\theta}{1-\beta_u \cos\theta}\right)^{2-\alpha}\simeq 1.2 \times10^{4} \] for $\alpha \simeq-0.5$, as expected nearer the core.

\section{Summary and Conclusions}
\label{sumconcl}

First analysis of the properties of component C80 were carried out by \citet{RocaSogorb:2010p11823}, concluding that although a helical shocked jet model -- including perhaps some bulk flow acceleration -- could explain the unusually large brightness temperature, it appears unlikely that it corresponds to the usual shock that moves from the core to the location of C80. Rather, \citet{RocaSogorb:2010p11823} proposed the need for an alternative process capable of explaining the high brightness temperature of C80, its appearance in high frequency images after April 2007, and its apparent stationarity. One of the possible alternatives considered by these authors involved a stationary shock at the location of C80 with emission suddenly enhanced by the arrival of a region of enhanced particle number in the jet flow, such as the one proposed to explain the kinematic and flux evolution properties of the HST-1 knot in \object{M87} \citep{Stawarz:2006p17247}.

The main goal of the new observations presented here is to further constrain the physical processes taking place in the jet of \object{3C~120} that have led to the extreme properties of C80. Thanks to the increased sensitivity achieved by our new observations we report on the existence of the A80 arc of total intensity and linear polarization associated to C80, as part of a larger bulge of emission that extends $\sim$20 mas along and across the jet. Most importantly, the polarization vectors are distributed perpendicularly to the semi-circular shape of A80, as it would be expected for the case of a compression by a shock front. This evidence, together with the excess in brightness temperature displayed by C80/A80 and its stationarity in flux and position support the model suggested by \citet{RocaSogorb:2010p11823} in which C80/A80 corresponds to a standing shock. More specifically, our modeling suggests a conical recollimation shock. Indeed, our simulations based on the synchrotron emission from a conical shock, as described by \citet{Cawthorne:2006p409}, reproduce quite closely the observed total and linearly polarized emission structure, the electric vector distribution, and the increased brightness temperature of C80/A80, allowing constraints on the values of the jet flow in \object{3C~120} and the geometry of the conical shock at $\sim80$\,mas from the core. In particular, our simulations provide the cone opening angle $\eta=10^{\circ}$, the jet viewing angle $\theta=16^{\circ}$ at the location of A80, and the upstream Lorentz factor $\gamma_u=8.4$.

An important issue to investigate further is the origin of the recollimation shock at such large distances from the core of emission in relativistic jets as in 3C~120, and presumably in M87. While for M87 it has been proposed that the recollimation originates from the transition between a parabolic to a conical shape \citep{Asada:2012p20001}, in 3C\,120 we suggest that the most plausible cause is a sudden drop in the external pressure, leading to the formation of a conical shock wave opening away from the nucleus.

The study of the nature of C80 -- and the structure of its related A80 arc of emission -- that we present in this paper were only possible thanks to the high angular resolution provided by the VLBA and its large sensitivity and good performance for polarimetric observations. The use of synthetic images of the total intensity and linear polarization of conical shocks has also proven to be a powerful tool to interpret the nature of jet structures, and to constrain the physical and geometrical properties of such structures and the jet plasma that forms them. Studies similar to that presented here can be carried out for a number of cases -- additional potential examples might be the HST-1 and K1 knots in \object{M87} \citep{Stawarz:2006p17247} and \object{3C~380} \citep{Papageorgiou:2006p271}. Performing these studies will be important to obtain relevant information on the relativistic jets and their surrounding medium.

\begin{acknowledgements}
This research has been supported by the Spanish Ministry of Economy and Competitiveness grant AYA2010-14844 and by the Regional Government of Analuc\'{i}a grant P09-FQM-4784. We thank Dr. P. A. Hughes of the University of Michigan for useful discussion on the physics of recollimation shocks. The VLBA is an instrument of the National Radio Astronomy Observatory, a facility of the National Science Foundation operated under cooperative agreement by Associated Universities, Inc. This research has made use of the MOJAVE database maintained by the MOJAVE team \citep{Lister:2009p5316}.
\end{acknowledgements}



\end{document}